\documentstyle[aps,prl,multicol,epsf]{revtex}
\begin{document}

\title{\bf Traveling Waves, Front Selection, and 
Exact Nontrivial Exponents in a Random Fragmentation Problem}
\author{P.~L.~Krapivsky$^{1,2}$ and Satya N. Majumdar$^{2,3}$}
\address{$^1$Center for Polymer Studies and
Department of Physics, Boston University, Boston, MA 02215, USA}
\address{$^2$CNRS, IRSAMC, Laboratoire de Physique Quantique,
Universite' Paul Sabatier, 31062 Toulouse, France}
\address{$^3$Tata Institute of Fundamental Research, Homi Bhabba Road,
Mumbai-400005, India}

\maketitle

\begin{abstract}

\noindent
We study a random bisection problem where an initial interval of
length $x$ is cut into two random fragments at the first stage, then
each of these two fragments is cut further, etc.  We compute the
probability $P_n(x)$ that at the $n^{\rm th}$ stage, each of $2^n$
fragments is shorter than 1.  We show that $P_n(x)$ approaches a
traveling wave form, and the front position $x_n$ increases as
$x_n\sim n^{\beta}{\rho}^n$ for large $n$.  We compute exactly the
exponents $\rho=1.261076\ldots$ and $\beta=0.453025\ldots$ as roots of
transcendental equations.  We also solve the $m-$section problem where
each interval is broken into $m$ fragments. In particular, the
generalized exponents grow as $\rho_m\approx m/(\ln m)$ and
$\beta_m\approx 3/(2\ln m)$ in the large $m$ limit.  Our approach
establishes an intriguing connection between extreme value statistics
and traveling wave propagation in the context of the fragmentation
problem.

\end{abstract}

\begin{multicols}{2}

The statistics of extremes plays an important role in various branches
of physics, statistics, and mathematics\cite{Gumbel,Galambos}.  For
example, in physics of disordered systems, the statistics of extremely
low energy states governs the thermodynamic behavior at low
temperatures\cite{BM}.  The extreme-value statistics is well
understood when random variables are {\em independent} and identically
distributed.  Much less is known when the random variables are
correlated. In the replica language, this class of problems
corresponds to full replica symmetry breaking\cite{BM}.  Thus it would
be important to derive exact results for extreme value statistics for
correlated random variables.

On the other hand, there is a wide variety of problems in physics,
chemistry and biology, that allows for solutions with propagating
traveling waves\cite{VS1}. The problem of front propagation into an
unstable state dates back to the pioneering studies\cite{Fisher,KPP}
motivated by gene spreading in a population.  Numerous traveling wave
solutions were found in combustion\cite{zel} and reaction diffusion
systems\cite{Murray}; they also appear in the mean-field theory of
directed polymers in random medium\cite{DS}, calculations of Lyapunov
exponents of random matrices\cite{CD,VB}, pattern
formation\cite{cross}, and many other problems\cite{van2}.  Usually in
all these problems, the traveling wave has an exponentially decaying
front that advances with a uniform velocity from a stable state to an
unstable state as time increases.  Out of a continuum of possible
allowed velocities of the front, a unique value is usually
selected. This velocity selection mechanism has been investigated by a
variety of methods\cite{VS1,KPP,van2,Bram,BD,van1}, and it was found
that for ``sufficiently steep'' initial conditions, the minimal
velocity is usually selected.

Is there a connection between these two sets of problems?  In this
paper, we give a positive answer to this question in the context of a
random fragmentation problem\cite{note}.  Our study indicates that
there is perhaps a very general and deep connection between the two
seemingly unrelated topics of extreme value statistics and propagating
wave solutions.  Besides, we give an exactly solvable example of
extreme value statistics for correlated random variables.

The random bisection problem can be formulated as follows.  An
interval of length $x$ is cut into two halves of lengths $x_1=rx$ and
$x_2=(1-r)x$, respectively, where $r$ is the random number chosen from
the uniform distribution over $[0,1]$.  Next, each of these two
fragments is cut again cut into two parts (with no correlations
between the cuttings).  After the $n^{\rm th}$ step, there are $2^n$
fragments whose lengths are correlated random variables (correlations
are dynamically generated).  Given the initial length $x$, what is the
probability $P_n(x,l)$ that each of the $2^n$ fragments is shorter
than $l$?  In other words, what is the probability that the longest of
these $2^n$ fragments would be shorter than $l$.  The $l$ dependence
is simple, $P_n(x,l)=P_n(x/l)$, so in the following we set $l=1$
without loss of generality.

Clearly, $P_n(x)\equiv 0$ for $x\geq 2^n$.  Moreover, there exists a
threshold $x_n$ such that for large $n$, $P_n(x)\to 1$ for $x\ll x_n$
and $P_n(x)\to 0$ for $x\gg x_n$. The structure is that of an
advancing front and the front position $x_n$ increases with $n$ as
${\rho}^n$ for large $n$.  This intriguing ``phase transition" was
noticed by several workers\cite{robson,pittel,mahm} and finally
rigorously proved by Devroye\cite{D1} who also computed exactly the
value of $\rho=1.261076\ldots$.  Recently, Hattori and Ochiai\cite{HO}
conjectured, based on numerical simulation, that the more precise
asymptotic behavior of $x_n$ reads
\begin{equation}
\label{ho}
x_n\sim n^{\beta}{\rho}^n,
\end{equation}
and found numerically that $\beta\approx 0.407$.

In this paper, we derive this asymptotic result by using methods of
statistical physics.  In particular, we re-derive the value of $\rho$
in a physically transparent way and compute the exponent $\beta$
exactly, which is a completely new result.  We show that
$\beta=0.453025\ldots$ is nontrivial and is given by the root of a
transcendental equation.  Our technique consists of employing a
scaling ansatz which reduces the problem to solving an equation that
admits traveling wave solutions.  We then select an appropriate
solution by employing a well known front selection principle.  Our
technique also allows us to obtain exact results for a multisection
problem where each interval is cut into $m$ random pieces at every
stage. Once again, the distribution $P_n(x)$ follows the zero-one law,
with the threshold value $x_n(m)\sim n^{\beta_m}[\rho_m]^n$.  As in
the $m=2$ case, we compute the exponents $\rho_m$ and $\beta_m$
exactly for arbitrary $m$.  For example, $\rho_3=1.499118\ldots$ and
$\beta_3=0.429815\ldots$, and $\rho_m\to m/{\ln m}$ and $\beta_m\to
3/(2\ln m)$ when $m\to\infty$.  We also show that the variance of the
front position $x_n(m)$ is always finite.

We first consider the $m=2$ case, i.e., the bisection problem. It is
easy to see that $P_n(x)$ satisfies the exact recursion relation,
\begin{equation}
P_{n+1}(x)={1\over x}\int_0^{x}dy\,P_n(y)P_n(x-y),
\label{recur}
\end{equation}
with the initial condition, $P_0(x)=\theta(1-x)$, where $\theta(x)$ is
the Heaviside step function.  The prefactor $1/x$ on the right hand
side of Eq.~(\ref{recur}) is the probability that the point where the
interval is cut into two pieces is chosen randomly from the interval
$[0,x]$.  Taking the Laplace transform $Q_n(s)=\int_0^{\infty}
dx\,e^{-sx}P_n(x)$ of Eq.~(\ref{recur}), we get,
\begin{equation}
{ {dQ_{n+1}(s)}\over {ds}}=-Q_n^2(s).
\label{Lap1}
\end{equation}  
Physically, one expects that as $n$ grows, $P_n(x)$ will be nonzero
for $x<x_n$ and then will rapidly decay to zero for $x>x_n$ where $x_n$
is a threshold. An appropriate definition of $x_n$ would be,
\begin{equation}
x_n=\int_0^{\infty}dx\, P_n(x).
\label{xn}
\end{equation}
Thus, it is natural to rewrite $P_n(x)$ in the scaling form,
$P_n(x)=f_n(x/x_n)$.  Thence, the Laplace transform reads
$Q_n(s)=x_nF_n(sx_n)$ with $F_n(z)=\int_0^{\infty}dy\,e^{-zy}
f_n(y)$. Substituting this form of $Q_n(s)$ into Eq.~(\ref{Lap1}), we
get
\begin{equation}
{dF_{n+1}(z)\over {dz}}= -\left({{x_n}\over {x_{n+1}}}\right)^2
\left[F_n\left({ {zx_n}\over {x_{n+1}}}\right)\right]^2.
\label{F1}
\end{equation}
Note that, by definition, $Q_n(0)=x_n$ implying $F_n(0)=1$ for all
$n$. Additionally, $f_n(0)=1$ leads to $F_n(0)\to z^{-1}$ as $z\to
\infty$.  For convenience, we make a further substitution,
$F_n(z)=(1-H_n(z))/z$, which recasts Eq.~(\ref{F1}) into
\begin{equation}
z\,{dH_{n+1}\over{dz}}= H_n^2\left({ {zx_n}\over {x_{n+1}} }\right)
-2H_n\left({ {zx_n}\over {x_{n+1}} }\right)+H_n(z),
\label{H0}
\end{equation}  
with the boundary conditions $H_n(0)=1$ and $H_n(z)\to 0$ as $z\to
\infty$.

We first focus on the computing of $\rho$ in the asymptotic relation
(\ref{ho}).  To determine $\rho$, we seek a `stationary', i.e.,
independent of $n$, solution of Eq.~(\ref{H0}).  Using $x_n\sim
n^{\beta}{\rho}^n$ and taking the $n\to \infty$ limit, we find that
the stationary solution satisfies
\begin{equation}
z{dH(z)\over{dz}}= H^2\left({z\over {\rho}}\right)
-2H\left({z\over {\rho}}\right)+H(z),
\label{H1}
\end{equation}
where $H(0)=1$ and $H(z)\to 0$ as $z\to \infty$. While we could not
solve this non-local and nonlinear differential equation, we can
determine the ``eigenvalue'' $\rho$ exactly through the asymptotic
analysis.  Indeed, in the large $z$ limit, $H(z)$ is small and thus
one can neglect the nonlinear term in Eq.~(\ref{H1}). The resulting
linear equation admits a power law solution, $H(z)=az^{-\mu}$, with
\begin{equation}
\rho=\left[{{1+\mu}\over 2}\right]^{{1\over {\mu}}}.
\label{r1}
\end{equation}   
Thus, a wide range of possible $\mu$'s is in principle allowed.
However, usually a particular value is selected depending on the
initial condition of the system. This is very similar to the problem
of velocity selection in a large class of problems with wave
propagation\cite{Murray,Bram} and it is well known that for a wide
class of initial conditions, the extremum value is chosen. In the
present case, the function on the right hand side of Eq.~(\ref{r1})
has a unique maximum at $\mu={\mu}^*$, which is a root of $\ln
({{1+\mu^*}\over 2})=\mu^*/(1+\mu^*)$. Though we have not proved
explicitly that the extremum value is indeed chosen, one can infer
this conclusion from the general principle of front selection.  Note
also that $\rho=\rho(\mu^*)=1.261070\ldots$ can be written as
$\rho=e^{\alpha}$ where $\alpha$ is a solution of $\alpha=\log(2\alpha
e)$, in agreement with the result of Ref.\cite{D1}.

The exponent $\beta$ characterizes the next to leading asymptotic
behavior of the front.  Therefore, to compute $\beta$ we need to
consider the full equation (\ref{H0}) rather than its $n\to \infty$
limit.  A sub-leading asymptotic behavior of traveling fronts was
originally analyzed by Bramson\cite{Bram} for a reaction-diffusion
equation, and recently investigated in\cite{BD,van1,van2}.  Here we
employ an approach of Ref.\cite{BD}.  For finite but large $n$, we
make the following scaling ansatz for the function $H_n(z)$,
\begin{equation}
H_n(z)\approx n^{\alpha}G\left({{\ln z}\over {n^{\alpha}}}\right)z^{-\mu^*},
\label{beta0}
\end{equation}
where the exponent $\alpha$ and the scaling function $G(y)$ are yet to
be determined. The scaling function $G(y)$ must vanish as $y\to
\infty$. Also, $G(y)\sim y$ as $y\to 0$ to ensure that for large $n$,
$H_n(z)\sim z^{-\mu^*}$ and is independent of $n$.  We substitute this
scaling ansatz in Eq.~(\ref{H0}) and use $x_n\sim n^{\beta}{\rho}^n$
where $\rho$ is already known exactly from Eq.~(\ref{r1}).  Using the
exact value of $\mu^*$, we find that different leading order terms are
comparable only with the special choice $\alpha=1/2$. In that case,
the scaling function $G(y)$ satisfies an ordinary differential
equation,
\begin{equation}
(\ln\rho)^2\, {{d^2G}\over {dy^2}} 
+ y{{dG}\over {dy}} +(2\beta \mu^*-1)G(y)=0,
\label{para}
\end{equation}
with the boundary conditions $G(y)\sim y $ for $y\to 0$ and
$G(y)\to 0$ as $y\to \infty$.  This therefore constitutes an eigenvalue
problem where $\beta$ is the required eigenvalue.  The exact solution
of this differential equation that satisfies the boundary condition at
$y\to \infty$ is given by\cite{GR},
\begin{equation}
G(y) = A\,e^{-{y^2\over {4\ln^2 \rho}}}\,
D_{2(\beta\mu^*-1)}\left( {y\over {\ln \rho}}\right) ,
\label{para2}
\end{equation}
where $A$ is a constant and $D_p(x)$ is the parabolic cylinder
function of order $p$ and argument $x$.  From the known properties of
cylinder functions\cite{GR}, we find that the boundary condition,
$G(y)\sim y$ as $y\to 0$, selects the eigenvalue
$2(\beta\mu^*-1)=1$. This is because only the parabolic cylinder
function with index 1 vanishes linearly with $x$ as $x\to 0$. Thus,
the exponent $\beta$ is exactly determined in terms of the known
$\mu^*$,
\begin{equation}
\beta= {3\over {2\mu^*}}.
\label{beta}
\end{equation} 
Using $\mu^*=3.311070\ldots$, we get $\beta=0.453025\ldots$.  Our
exact result slightly differs from the numerical value $\beta\approx
0.407$ obtained in Ref.\cite{HO}.  An accurate numerical determination
of $\beta$ is not simple as it is a sub-leading correction to the the
leading asymptotic behavior.
 
The bisection problem can be straightforwardly generalized to the
$m$-section problem where at each stage, every interval is cut into
$m$ random pieces\cite{D2}. The probability $P_n(x)$ that at the
$n^{\rm th}$ stage each of $m^n$ fragments is shorter than 1
satisfies the exact recursion relation,
\begin{eqnarray}
\label{recurm}
P_{n+1}(x)={ {(m-1)!}\over {x^{m-1}}}
\int_0^\infty &\dots& \int_0^{\infty}
\prod_{j=1}^{m} dy_j\,P_n(y_j)\nonumber\\
&\times&\delta\left(\sum_{i=1}^m y_i-x\right),
\end{eqnarray}
with the initial condition, $P_0(x)=\theta(1-x)$. This equation can be
easily derived as follows. At a given stage $n$, an interval of length
$x$ is cut into $m$ fragments. Let $z_1,\ldots, z_{m-1}$ denote the
location of the points at which the interval is cut. The allowed range
of values of the co-ordinates $[z_1,\ldots, z_{m-1}]$ is the
$(m-1)$-dimensional simplex: $x\geq z_{m-1}\geq \ldots\geq z_1 \geq
0$.  The volume of this simplex is $x^{m-1}/(m-1)!$, and this explains
the prefactor of Eq.~(\ref{recurm}). Finally, by changing co-ordinates
to $y_1=z_1, y_2=z_2-z_1, \ldots, y_m=x-z_{m-1}$, one obtains
Eq.~(\ref{recurm}).  Note also that for $m=2$, Eq.~(\ref{recurm})
reduces to Eq.~(\ref{recur}).  We now proceed exactly as in the $m=2$
case. The Laplace transform, $Q_n(s)$, satisfies the differential
equation,
\begin{equation}
{ {d^{m-1}Q_{n+1}(s)}\over {ds^{m-1}}}
=(-1)^{m-1}(m-1)!\left[Q_n(s)\right]^m.
\label{Lap2}
\end{equation}
We again expect the threshold to grow as $n^{\beta_m}(\rho_m)^n$. To
determine $\rho_m$, we assume that $P_n$ approaches the scaling form,
$P_n(x)=f(x/x_n(m))$, in the large $n$ limit. 
Thence, $Q_n(s)=x_n(m)F(sx_n(m))$ where
$x_n(m)=\int_0^{\infty}dx\,P_n(x)$ and
$F(z)=\int_0^{\infty}dz\,e^{-zy}f(y)$.  By inserting the scaling form
for $Q_n(s)$ and $x_n(m)\sim n^{\beta_m}(\rho_m)^n$ into
Eq.~(\ref{Lap2}) we finally arrive at the non-local differential
equation,
\begin{equation}
{{ d^{m-1}F(z)}\over {dz^{m-1}} }
={ {(-1)^{m-1}(m-1)!}\over {(\rho_m)^m}} 
\left[F\left({z\over \rho_m}\right)\right]^m.
\label{F2}
\end{equation}
Substituting $F=(1-H(z))/z$ and linearizing the resulting equation for
$H(z)$ in the large $z$ limit, one finds a solution that behaves
algebraically, $H(z)\sim z^{-\mu_m}$, in the large $z$ limit.  A
straightforward algebra then shows that $\rho_m$ depends on $\mu_m$ via 
\begin{equation}
\rho_m=  \left[{{\Gamma(\mu_m+m)}\over 
{\Gamma(\mu_m+1)\Gamma(m+1)}}\right]^{{1\over \mu_m}},
\label{r2}
\end{equation}
where $\Gamma(x)$ is the usual Gamma function.  Equation (\ref{r2}) is
the generalization of Eq.~(\ref{r1}) for the arbitrary $m$ case. Once
again, the function on the right hand side has a unique maximum at
$\mu_m=\mu_m^*$ and this maximum is actually selected.  In particular,
$\rho_3=\exp[ {{(2b+3)}\over {(b+1)(b+2)}}]$ where $b$ is found from
$b(2b+3)=(b+1)(b+2)\ln [(b+1)(b+2)/6]$.  Solving this numerically,
gives $\rho_3=1.499118\ldots$ and $\mu_3^*=3.489870\ldots$.

The exponent $\beta_m$ can be determined exactly for arbitrary $m$.
We do not repeat the calculation as it follows the same steps as for
$m=2$ and the final result is given by the same expression
(\ref{beta}), i.e., $\beta_m=3/(2\mu_m^*)$.  For instance,
$\beta_3=0.429815\ldots$, and generally the exponent $\beta_m$
decreases with increasing $m$.

One can easily derive the asymptotic behavior of $\rho_m$ and
$\beta_m$ for large $m$. Taking logarithms on both sides of
Eq. (\ref{r2}) and differentiating with respect to $\mu_m$ we find,
after straightforward algebra, that the maximum occurs at
$\mu_m^*\approx \ln m$. In this calculation, we have used properties
of Gamma function for large arguments: $\Gamma(m+\mu_m)/
\Gamma(m+1)\sim (m+1)^{\mu_m-1}$ for $m\gg \mu_m$. Substituting
$\mu_m=\ln m$ into Eqs.~(\ref{r2}) and (\ref{beta}), we find that to
leading order for large $m$,
\begin{equation}
\rho_m\approx {m\over {\ln m}}\quad {\rm and}\quad 
\beta_m\approx {3\over {2\ln m}}.
\label{log1}
\end{equation}

In this work, we have established a relationship between two seemingly
disparate subjects -- the statistics of extremes and traveling wave
propagation.  Specifically, we have shown that the probability density
of the maximal fragment length, which is the derivative of the
distribution $P_n(x)$, approaches the solitary traveling wave in the
large $n$ limit.  This traveling wave has a finite width which
implies, in the context of the random binary search trees, that the
variation of the height of a tree is finite.  This result has long
been anticipated on the basis of numerical experiments but remained
intractable\cite{mahm,D3}.

The present work admits several extensions. One could assume that an
interval can be cut into fragments of relative lengths $r$ and $1-r$
with the probability density $\pi(r)$ which is arbitrary apart from
normalization and symmetry requirements, $\pi(r)=\pi(1-r)$ and
$\int_0^1dr\,\pi(r)=1$.  Besides the uniform probability density,
$\pi(r)=1$, one could solve the random bisection problem for a number
of other densities, e.g., for $\pi(r)=6r(1-r)$.  The basic result,
Eq.~(\ref{ho}), always holds but the exponents do depend on the
probability density $\pi(r)$.

One could also modify the random bisection problem by deciding to cut
only those intervals which are still longer than 1. This problem can
be solved by employing similar techniques as the original one.  The
most interesting question is how does the total number of intervals
$N(x)$, which are left after all intervals will be shorter than 1,
depend on the initial length $x$.  For the original random bisection
problem with the uniform probability density, our solution implies
$N(x)\sim x^\gamma (\ln x)^{-\delta}$ with $\gamma=\ln 2/\ln \rho\cong
2.9881$ and $\delta=\beta\gamma\cong 1.35368$.  For the modified
version, we have found $N(x)\simeq 2x$.  Thus, the modified algorithm
is much more effective if we want to minimize the total number of
cuts.

Apart from the straightforward applications of our results to the random
search tree problem in computer science, our results have possible implications
in a number of topics of
current interest in physics and chemistry. One obvious application is
to the fracture of a rock, or to the beak-up of a polymer. Another possible
application is to granular materials. Recent experiments have studied the 
propagation of stresses in a granular pile of glass beads subjected to a
large vertical overload\cite{Nagel}. Our model is closer to the situation when
the vertical overload is localized. This force gets transmitted from the top layer 
to the bottom layer. 
If this external force greatly exceeds the weight of individual
grains, $F\gg w$, and if fractions of force transmitted from a grain
to its neighbours in the lower layer are random, then the model studied above
gives the mean-field description of the force transmission. Within this approximation,
the force chains starting from the top never intersect with each other thus maintaining
the tree structure. Our results then imply that
if the force from a grain in a given layer is always transmitted to $m$ grains in the next
layer, than the granular material should contain at least
$\ln(F/F_*)/\ln\rho_m$ layers to guarantee that the normal force supported by any grain at the bottom layer
never exceeds $F_*$.

\medskip
\noindent
PLK acknowledges support from NSF, ARO, and CNRS.

\end{multicols}

\begin{references}
\bibitem{note}    An equivalent version of the random bisection problem
                  has been originally studied by computer scientists in 
                  the context of random binary search 
                  trees\cite{robson,pittel,mahm}.
\end{references}
\end{document}